\documentclass[twocolumn]{jpsj3}

\usepackage{times}
\usepackage{graphicx}

\title{Heavy-Electron Formation and Polaron-Bipolaron Transition
in the Anharmonic Holstein Model}

\author{Takahiro \textsc{Fuse}$^1$, Yoshiaki \textsc{\=Ono}$^2$,
and Takashi \textsc{Hotta}$^{1,3}$}

\inst{$^1$Department of Physics, Tokyo Metropolitan University,
Hachioji, Tokyo 192-0397, Japan \\
$^2$Department of Physics, Niigata University, Ikarashi,
Niigata 950-2181, Japan \\
$^3$Advanced Science Research Center, Japan Atomic Energy Agency,
Tokai, Ibaraki 319-1195, Japan}

\recdate{\today}

\abst{
The emergence of the bipolaronic phase and the formation of the heavy-electron state
in the anharmonic Holstein model are investigated using the dynamical mean-field theory 
in combination with the exact diagonalization method.
For a weak anharmonicity,
it is confirmed that the first-order polaron-bipolaron transition occurs from
the observation of a discontinuity in the behavior of several physical quantities.
When the anharmonicity is gradually increased, the polaron-bipolaron transition
temperature is reduced as well as the critical values 
of the electron-phonon coupling constant for polaron-bipolaron transition. 
For a strong anharmonicity, the polaron-bipolaron transition eventually changes to a crossover behavior.
The effect of anharmonicity on the formation of the heavy-electron state 
near the polaron-bipolaron transition and the crossover region is discussed in detail.
}

\kword{Heavy electron, anharmonicity, rattling, bipolaron,
dynamical mean-field theory}

\begin{document}
\maketitle

\section{Introduction}

Recently, heavy-electron phenomena have attracted renewed attention
in the research field of condensed matter physics.\cite{ICHE2010} 
A traditional mechanism of the emergence of the heavy-electron state
is based on quantum criticality induced by the competition
between the Kondo effect and Ruderman-Kittel-Kasuya-Yosida interaction.
The Kondo effect due to local magnetic moment has been well understood,
\cite{Kondo40} but the Kondo-like phenomenon occurs in a more general manner,
when a localized entity with internal degrees of freedom is embedded
in a conduction electron system and quantum-mechanical exchange
interaction effectively works between local degrees of freedom and
conduction electrons.

In particular, the Kondo phenomenon with a phonon origin has been potentially discussed. 
First, Kondo himself has considered a conduction electron system 
coupled with a local double-well potential.\cite{Kondo1,Kondo2}
Having two possible electron position in a double-well potential
play roles in pseudo-spins, and the Kondo-like behavior
is considered to appear in such a two-level system.
In fact, it has been shown that the two-level Kondo system exhibits
the same behavior as the magnetic Kondo effect.
\cite{Vladar1,Vladar2,Vladar3} 
Another important issue regarding the Kondo effect with a phonon origin
has been shown by Yu and Anderson.\cite{Yu-Anderson}
They have pointed out that the scattering process between spinless
$s$-wave conduction electrons to $p$-wave ones is induced by ion displacement.
The model proposed by Yu and Anderson has been shown to be mapped
to the two-level Kondo model
at low temperatures.\cite{Matsuura1,Matsuura2}

A recent revival of research on the Kondo effect with a phonon origin
has been triggered by active experimental investigations
on cage structure compounds such as
filled skutterudites,\cite{SkutReview,GotoSkut,KotegawaSkut,Sanada}
clathrates,
\cite{SalesClat,GotoClat,AvilaClat1,AvilaClat2,NemotoClat1,NemotoClat2}
and $\beta$-pyrochlore oxides.
\cite{HiroiGrp_bp1,HiroiGrp_bp2,HiroiGrp_bp3,HiroiGrp_bp4,Umeo_bp}
In these materials, a guest atom is contained in a cage composed of
relatively light atoms and oscillates with a large amplitude
in a potential with a strong anharmonicity.
Such local oscillation with a large amplitude is called {\it rattling}, and
exotic phenomena induced by rattling have attracted much attention
in the research of strongly correlated electron materials with
a cage structure.

An example of such active investigations is
the study of a magnetically robust heavy-electron behavior observed
in SmOs$_4$Sb$_{12}$.\cite{Sanada}
The electronic specific heat coefficient $\gamma_{\rm e}$ is
in proportion to the effective mass of electrons,
but in this material,
$\gamma_{\rm e}$ is several hundred times larger than that of
a free-electron system and is found to be almost unchanged
even if a magnetic field up to 30 Tesla is applied.
If the heavy effective mass originates from the quantum criticality
due to the traditional Kondo effect with a magnetic origin,
$\gamma_{\rm e}$ should be significantly suppressed
by the applied magnetic field.
The origin of the heavy-electron state in SmOs$_4$Sb$_{12}$ is
considered to be electron-rattling interaction. 
In fact, four- and six-level Kondo systems have been analyzed
and the magnetically robust heavy-electron state has been actually
obtained.\cite{Yotsuhashi,Hattori1,Hattori2}
The periodic Anderson-Holstein model has been analyzed using 
the dynamical mean-field theory, 
and the mass enhancement due to large lattice fluctuations
and phonon softening towards a double-well potential
have been addressed.\cite{Mitsumoto0,Mitsumoto1,Mitsumoto2}

Kondo phenomena in the conduction electron system coupled with
local Jahn-Teller phonons~\cite{Hotta1,Hotta2}
and Holstein phonons~\cite{Hotta3,Hotta4}
have been discussed for the realization of a nonmagnetic Kondo effect. 
From the numerical evaluation of $\gamma_{\rm e}$
in the conduction electron system coupled with
local anharmonic Holstein phonons,
it has been shown that $\gamma_{\rm e}$ is largely enhanced by rattling
and is actually robust under an applied magnetic field.\cite{Hotta5}
Furthermore, it has been pointed out that the Kondo effect
due to rattling should exhibit a peculiar isotope effect,
which is experimental evidence of rattling-induced
heavy-electron phenomena.\cite{Hotta6}
Quite recently, the vibration of magnetic ions has been explicitly
included in the spinful Yu-Anderson model, and the two-channel Kondo
effect has been comprehensively discussed.\cite{Yashiki1,Yashiki2,Yashiki3}

When we turn our attention to the $\beta$-pyrochlore oxide KOs$_2$O$_6$,
significant effects of the rattling of a K ion in a tetrahedral cage
composed of O and Os ions have been discussed.
The rattling-associated anomaly is found in the form of
the first-order structural transition at $T_p \sim 7.5$ K,
which is difficult to be influenced by a magnetic field
and is not accompanied by any symmetry breaking.
\cite{HiroiGrp_bp3,HiroiGrp_bp4}
It has been reported that electric resistivity is
in proportion to $T^2$ for $T<T_p$,
while it behaves as $\propto \sqrt{T}$ for $T>T_p$.
In the peculiar behavior at high temperatures,
it has been suggested that anharmonic phonons play important roles.
\cite{Dahm-Ueda}
The transition at $T_p$ has been discussed in the context of
electron-rattling interaction.
A possible liquid-gas-type rattling transition and the multipole transition
driven by the octupole component of K ion rattling have been pointed out.
\cite{HatryTntg1,HatryTntg2} From the analysis of
the harmonic Holstein model,
it has been reported that the first-order transition from the polaronic
state to the bipolaronic state occurs in the strong-interaction region
at low temperatures.\cite{Peeters1,Peeters2,Peeters3,FuseOno}

In this study, to obtain deep insight into the effect of
anharmonicity on the polaron-bipolaron transition and the heavy-electron state,
we analyze the anharmonic Holstein model
using the dynamical mean-field theory (DMFT) in combination with
the exact diagonalization method at low temperatures such as
$10^{-4}$ of the conduction electron bandwidth.\cite{FuseOno2}
When the anharmonicity is weak, i.e., for nearly harmonic phonons,
we again find the first-order polaron-bipolaron transition due to the
observation of a discontinuity in the behavior of physical quantities.
When the anharmonicity is increased, the polaron-bipolaron transition
temperature is reduced and the critical value of
electron-phonon interaction becomes smaller.
For a strong anharmonicity, the polaron-bipolaron transition disappears
and it becomes to a crossover behavior.
We discuss the effect of anharmonicity on
the heavy-electron state near the polaron-bipolaron transition
and crossover regions.

The organization of this paper is as follows.
In \S ~2, we introduce the anharmonic Holstein model and
explain the anharmonic potential used in this paper.
We also provide a brief explanation of the DMFT. 
In \S ~3, we discuss the properties of the system obtained using the DMFT.
Several physical quantities are discussed when we change
the electron-phonon interaction, the anharmonicity of phonons, and temperature. 
Finally, in \S ~4, we summarize this paper.
Throughout this paper, we use such units as $\hbar=k_{\rm B}=1$.

\section{Model and Method}

In this section, we introduce the Hamiltonian as
\begin{equation}
 H=\sum_{\mib{k}\sigma}
  \varepsilon_{\mib{k}}c_{\mib{k}\sigma}^{\dag}c_{\mib{k}\sigma}+H_{\rm eph},
\end{equation}
where $\varepsilon_{\mib{k}}$ is the energy of conduction electrons,
$c^\dag_{\mib{k}\sigma}$ is a creation operator for 
conduction electrons with a wave vector $\mib{k}$ and a spin $\sigma$,
and $H_{\rm eph}$ denotes a local electron-phonon term.
In the following, we describe $H_{\rm eph}$ and the potential
for the vibration of the guest atom.
Then, we briefly explain the scheme of the DMFT to solve the Hamiltonian.

\subsection{Electron-phonon coupling term}

We consider a situation in which electrons are coupled with
the local oscillation of an atom.
Such a situation is expressed by
\begin{equation}
  \label{eq:Hloc}
  H_{\rm eph}= \sum_{i} \left[ gQ_{i} (n_i-1)
           +\frac{P_{i}^2}{2M}+V(Q_i) \right],
\end{equation}
where $g$ denotes the coupling constant between electron density and
the oscillation of the atom,
$i$ indicates the atomic site,
$n_i=\sum_{\sigma}c^\dag_{i \sigma}c_{i \sigma}$,
$c_{i \sigma}$ is the annihilation operator of an electron
with a spin $\sigma$ at a site $i$,
$Q_{i}$ is the normal coordinate of breathing-mode oscillation of the atom,
$P_{i}$ indicates the corresponding canonical momentum,
$M$ is mass of oscillator atom,
and $V(Q_i)$ is a potential for atom, which is 
expressed by
\begin{equation}
 V(Q_{i}) = k_2 Q_{i}^2 +k_4 Q_{i}^4.
\end{equation}
Here, $k_2$ and $k_4$ respectively denote the coefficients for
the second- and fourth-order terms of the potential for the atom.

Note that in the coupling between electrons and oscillation,
we subtract unity from the electron number
for convenience in the numerical calculation.
If we do not carry out this subtraction, the electron-phonon coupling
is effectively enhanced at doubly occupied sites
and it is necessary to prepare a large number
of phonon basis to obtain convergent results.
Since we are interested in the bipolaronic state,
it is crucial to perform the calculation with a significant amount of 
double occupancy with sufficient precision.
Thus, we use the electron-phonon term in eq.~(\ref{eq:Hloc}).

For more calculations, it is convenient to introduce
the second-quantized phonon operator by following the standard
quantum mechanical procedure.
The displacement is expressed by $Q_{i} = Q_0 (b_{i} + b_{i}^{\dagger})$,
where $Q_0=1/\sqrt{2M\omega_0}$, $\omega_0$ is the phonon energy,
and $b_i$ is the annihilation operator of phonons.
Note that $Q_0$ denotes the amplitude of zero-point
oscillation for harmonic phonons.
Since we consider the anharmonic oscillation
including the case of a negative $k_2$,
we do not impose an explicit relation between
$k_2$ and $\omega_0$.

Then, we rewrite $H_{\rm eph}$ using phonon operators as
\begin{equation}
 \begin{split}
  H_{\rm eph} &= \omega_0 \sum_{i}
             \Bigl[ \sqrt{\alpha}(b_i+b_i^{\dag}) (n_{i}-1)
           +b_i^{\dag}b_i+1/2 \\
           &+\beta_2(b_i+b_i^{\dag})^2
           +\beta_4 (b_i+b_i^{\dag})^4 \Bigr],
 \end{split}
\end{equation}
where $\alpha$, $\beta_2$, and $\beta_4$ are given by
\begin{equation}
  \alpha=\frac{g^2}{2M\omega_0^3},~
  \beta_2=\frac{1}{4}\Bigl( \frac{2k_2}{M\omega_0^2}-1 \Bigr),~
  \beta_4=\frac{k_4}{4M^2\omega_0^3}.
\end{equation}
Among them, $\alpha$ indicates the nondimensional electron-phonon
coupling constant, while $\beta_2$ and $\beta_4$ denote
the nondimensional second- and fourth-order anharmonicity parameters, respectively.
Note that the anharmonicity of the potential
is controlled by adjusting $\beta_2$, which becomes both positive and negative,
while we always set $\beta_4>0$ to
restrict the oscillation of an atom in a finite space.
We also note that the harmonic case is denoted by $\beta_2=\beta_4=0$.

\subsection{Anharmonic potential}

Now, we discuss the anharmonic potential for the oscillation of an atom.
For this purpose, it is convenient to introduce the
nondimensional displacement $q_i$ through the relation 
$q_i=Q_i / Q_0$ in accordance with the second-quantization of a phonon.
Using nondimensional displacement, we obtain the potential as
\begin{equation}
  V(q_i)=\omega_0 \left[\left(\beta_2 +\frac{1}{4}\right)q_i^2
        +\beta_4 q_i^4 \right].
  \label{eq:br_loc_pot}
\end{equation}
Note that the energy scale of $V(q_i)$ is taken as $\omega_0$.

\begin{figure}[t]
\begin{center}\leavevmode
\rotatebox{0}{\includegraphics[width=75mm,angle=0]{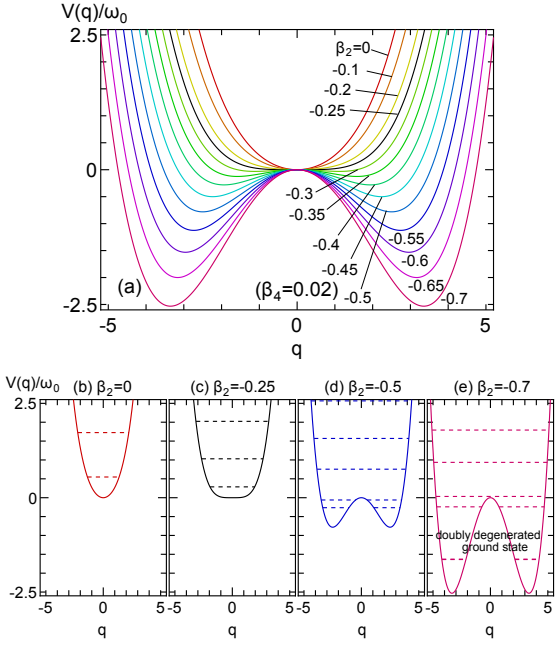}}
\caption{(Color online)
(a) Local phonon potential $V(q)$ for several values of $\beta_2$
from $0$ to $-0.7$ with $\beta_4=0.02$. 
(b)-(e) Several typical potentials with $\beta_2=0$, $-0.25$, $-0.5$, and $-0.7$, respectively.
We show eigenenergy levels with dotted horizontal lines.
\label{fig:potentialshapes}
}
\end{center}
\end{figure}

In Fig.~\ref{fig:potentialshapes}(a),
we show several anharmonic potentials by changing
$\beta_2$ for $\beta_4=0.02$.
For $\beta_2>-0.25$,
the potential minimum is always situated at $q_i=0$ and
the shape of a single-well potential is similar to
that for a harmonic potential.
However, in the region of $0>\beta_2>-0.25$,
the bottom of the potential is found to be wide and flat,
since at $\beta_2=-0.25$, the second-order term
of the potential vanishes.
When $\beta_2$ is decreased from $-0.25$,
the potential shape is changed,
since potential minima appear at $\pm q_{\rm min}$,
given by $q_{\rm min}=\sqrt{(\beta_2+1/4)/(2\beta_4)}$.
For $\beta_2<-0.25$, the potential minimum rapidly decrease, 
leading to the double-well potential.
We consider that the decrease in $\beta_2$ indicates
the increase in anharmonicity.

In Figs.~\ref{fig:potentialshapes}(b)-\ref{fig:potentialshapes}(e),
we show some typical shapes for a single-well type for $\beta_2=0$,
a flat-bottom type for $\beta_2=-0.25$,
a shallow double-well type for $\beta_2=-0.5$,
and a deep double-well type for $\beta_2=-0.7$.
We also show the eigenenergy levels obtained by
the diagonalization of $H_{\rm eph}$.
It is observed that the energy levels tend to move to the lower side
and that the width between adjacent levels becomes smaller,
when $\beta_2$ is decreased.
In the double-well type cases shown in
Figs.~\ref{fig:potentialshapes}(d) and \ref{fig:potentialshapes}(e),
several levels are found inside the wells. 
When we further decrease $\beta_2$,
the levels of the ground and first-excited states exist deep 
inside the wells
and the width between adjacent levels is extremely small.
We expect that the ground and first-excited states finally
become degenerate for a large negative $\beta_2$
with a deep double-well type potential.

\begin{figure}[t]
\begin{center}\leavevmode
\rotatebox{0}{\includegraphics[width=60mm,angle=0]{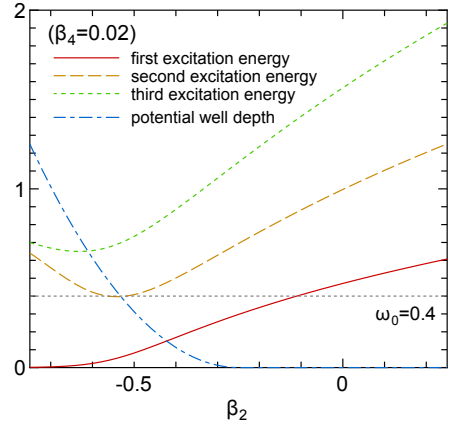}}
\caption{(Color online) 
 Energies in the local phonon potentials.
 The first, second, and third excitation energies of the local phonon
 and the depth of the potential well are shown.
 The phonon frequency $\omega_0=0.4$ is also indicated.
 \label{fig:energylevels}
}
\end{center}
\end{figure}

In Fig.~\ref{fig:energylevels}, 
we show several excitation energies
and the depth of the potential well,
defined by $V(0)-V(q_{\rm min})$.
We also depict $\omega_0=0.4$ line in the graph, 
which corresponds to the first excitation energy
in the harmonic phonon case.
When we decrease $\beta_2$, i.e., increase anharmonicity,
we find that the excitation energies are totally suppressed.
For $\beta_2<-0.25$, the potential wells are formed
at $q_i=\pm q_{\rm min}$.
When we decrease $\beta_2$, the first excitation energy is gradually
decreased and eventually at $\beta_2 \sim -0.7$, it becomes almost zero.
On the other hand, the second and third excitation energies are rather
increased in the region of $\beta_2<-0.5$,
since the ground-state energy is rapidly decreased.

\subsection{Dynamical mean-field theory}

To solve the model Hamiltonian,
it is necessary to use appropriate approximation depending on the problem.
Here, we adopt the dynamical mean-field theory (DMFT).\cite{A.Georges}
In the present case, the model is mapped onto an effective impurity
Anderson-Holstein model.\cite{A.C.Hewson}
In the following, we briefly explain the scheme of the DMFT.

The local electron Green's function $G(i\omega_n)$ is given by 
\begin{equation}
   G(i\omega_n) = \sum_{\mib{k}}
   \frac{1}{i\omega_n-\varepsilon_{\mib{k}}-\mu-\Sigma(i\omega_n)},
\end{equation}
where $\omega_n$ is the fermion Matsubara frequency, given by
$\omega_n = \pi T (2n+1)$ with an integer $n$,
$T$ is the temperature,
$\mu$ is the chemical potential,
and $\Sigma(i\omega_n)$ is the electron self-energy.
Note that, in the DMFT, the momentum dependence of the self-energy
can be ignored.

To perform momentum summation,
it is necessary to specify the lattice type.
Here, we consider the Bethe lattice with a semielliptic
density of states (DOS) given by
\begin{equation}
  \rho(\varepsilon)=\frac{4}{\pi W^2}\sqrt{W^2-4\varepsilon^2},
\end{equation}
where $W$ is the bandwidth.
Using this DOS, we obtain the condition
for the local Green's function as
\begin{equation}
  G_0(i\omega_n)^{-1} = i\omega_n-\mu-(W/4)^2 G(i\omega_n),
\end{equation}
where $G_0(i\omega_n)$ is the bare local electron Green's function.
In the present work, we determine $G_0$ by the effective
impurity Anderson-Holstein model with $\alpha=0$ in an effective medium
determined self-consistently.
The effective impurity Anderson-Holstein model with a finite $\alpha$
is solved by the exact diagonalization method
for a finite-sized cluster to obtain $G(i\omega_n)$ at a finite temperature,
given by
\begin{eqnarray}
 \label{Gel}
 G(i\omega_n) =
  \frac{1}{Z} \sum_{j,\ell}
  \frac{e^{-E_{\ell}/T}+e^{-E_j/T}}
       {i\omega_n +E_{j}-E_{\ell}}
 \left|\langle j | c_{i\sigma} | \ell \rangle\right|^2,
\end{eqnarray}
where $Z$ is the partition function given by
$Z=\sum_j e^{-E_j/T}$,
$|j\rangle$ is the eigenstate of the effective impurity
Anderson-Holstein model,
and $E_j$ is the corresponding eigenenergy.

In the present paper,
we use the five-site cluster and the cutoff number $N_{\rm c}$ for
the phonon basis is set to be $12$. 
We have checked the convergence of the calculations
in comparison with the results
for the six-site cluster and $N_{\rm c}=15$. 
We concentrate our attention to the half-filling case
with $\langle n_i \rangle=1$. 
We set $W=4$ and $\omega_0=0.4$ in the following calculations.
Note that we restrict ourselves to the case with the normal state 
without any symmetry breaking.

\section{Results of DMFT Calculations}

\subsection{Lattice fluctuations}

\begin{figure}[t]
\begin{center}\leavevmode
\rotatebox{0}{\includegraphics[width=70mm,angle=0]{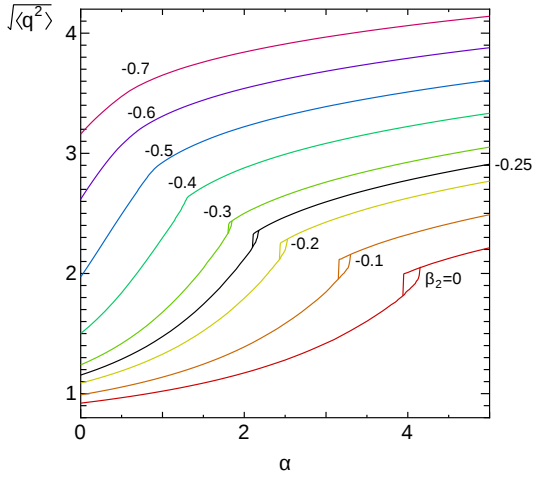}}
\caption{(Color online) 
 Lattice fluctuations $\sqrt{\langle{q^2}\rangle}$ vs $\alpha$
 for several $\beta_2$ values with $\beta_4=0.02$ and $T=0.0025$.
 \label{fig:alp-rq2_t00025}
}
\end{center}
\end{figure}

Let us start our discussion about the anharmonicity dependences
of various physical quantities for a fixed temperature.
First, we discuss the anharmonicity effect on the lattice fluctuation
$\sqrt{\langle{q^2}\rangle}$.
In Fig.~\ref{fig:alp-rq2_t00025},
we show the $\alpha$ dependence of the lattice fluctuation
$\sqrt{\langle{q^2}\rangle}$ 
for several values of $\beta_2$ with $\beta_4=0.02$ and $T=0.0025$. 
It is found that the curves for $\sqrt{\langle{q^2}\rangle}$
are monotonically increasing functions.
In the following, we discuss
the change in the curves for different values of $\beta_2$.

When we decrease $\beta_2$, i.e., strengthen the anharmonicity,
the amplitude of the lattice fluctuations becomes larger.
In fact, at $\alpha=5$, the lattice fluctuation increases from $2.2$
to $\sim 4.1$ when $\beta_2$ decreases from $0$ to $-0.7$.
Note that the lattice fluctuation at $\alpha=0$ already exhibits
a similar tendency as it increases from $\sim 0.9$ to $\sim 3.2$
when $\beta_2$ is changed from $0$ to $-0.7$.
This tendency can be understood from the potential shape,
as shown in Fig.~1.
Namely, when $\beta_2$ is decreased, the total width of the
potential becomes larger and the amplitude of the oscillation
in the potential increases owing to thermally activated
and quantum tunneling motions.

Next, we comment on the change in the function shape due to $\beta_2$.
For small absolute values of $\beta_2$,
the shape of the curve is convex-downward for a small $\alpha$, while
it becomes slightly convex-upward for a large $\alpha$.
For instance, for $\beta_2=0$, the change in the behavior
can be observed at $\alpha \sim 4$.
When the absolute value of $\beta_2$ becomes larger, 
the curve becomes steeper and $\alpha$ at which
the convex-downward behavior changes to a slightly convex-upward one becomes smaller.
Finally, for $\beta_2= -0.6$ or $-0.7$, 
the behavior for a small $\alpha$ also becomes a linear function.
These changes are thought to be caused by the enhancement of
the effective electron-phonon interaction coupled with
the anharmonic potential.

For a negative $\beta_2$ with small absolute value, 
we find the hysteresis region at approximately $\alpha$ 
where the increasing behavior is changed. 
In actual calculations, we obtain two solutions for the same 
$\alpha$ by gradually increasing and decreasing $\alpha$. 
One of the solutions is stable and the other is metastable. 
Thus, this region can be accounted for by the hysteresis behavior 
observed in the experimental measurements 
as well as by the first-order phase transition point 
where the stable and metastable solutions are switched.

With decreasing $\beta_2$, the hysteresis region becomes smaller.
For $\beta_2 <-0.5$, it eventually disappears and becomes a smooth
crossover between the two solutions.
Since the size of the hysteresis region is considered to be related
to the energy scale of the first-order phase transition temperature,
this result indicates that the anharmonicity suppresses
the first-order phase transition.
Note that in the present calculations,
the systematic reduction in the size of the hysteresis region
is obtained at a certain accuracy.
When we attempt to quantitatively obtain improved results,
it is necessary to resort to an extrapolation technique
to estimate the size of the hysteresis region.

\begin{figure}[t]
\begin{center}\leavevmode
\rotatebox{0}{\includegraphics[width=65mm,angle=0]{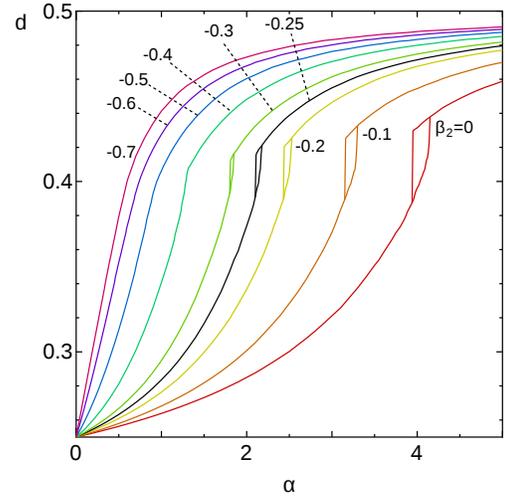}}
\caption{(Color online) 
 Double occupancy $d$ vs $\alpha$
 for several $\beta_2$ values with $\beta_4=0.02$ and $T=0.0025$.
 \label{fig:alp-doc_t00025}
}
\end{center}
\end{figure}

\subsection{Double occupancy}

Now we move to the discussion of the electronic states.
For this purpose, we examine the behavior of the double occupancy
$d=\langle{n_\uparrow n_\downarrow}\rangle$
as one of the typical physical quantities.
In Fig.~\ref{fig:alp-doc_t00025}, $d$ is plotted as a function
of $\alpha$ for several $\beta_2$ values with $\beta_4=0.02$
and $T=0.0025$.
For $\alpha=0$, we find $d=0.25$, since four local electron states
are degenerate in the noninteracting case.
When we increase $\alpha$, 
$d$ increases monotonically, but the increasing behavior changes
during the transition or the crossover.
For a small $\alpha$, $d$ is a convex-downward (almost linear) function
at a large (small) $\beta_2$, 
while it is a convex-upward function for a large $\alpha$.
For a sufficiently large $\alpha$, $d$ asymptotically approaches $0.5$, 
since the vacant and the double-occupied states are degenerate at
half-filling in the large $\alpha$ limit.

We remark that $d$ directly indicates the amount of the bipolaronic state
in which electrons with antiparallel spins are coupled
due to the effective attractive interaction mediated by phonons.
Thus, we characterize the two solutions in the hysteresis region using $d$.
Namely, among the two solutions, we define the solution
with a larger $d$ as {\it the bipolaronic solution}, while
that with a smaller $d$ is called {\it the polaronic solution}.
Note that, in the definition,
the polaronic solution does $not$ indicate $d=0$.

When we consider the $\beta_2$ dependence, we again find several remarks
that we have already described in the previous subsection.
Namely, the coexistence region disappears for a small $\beta_2$ 
and the polaron-bipolaron transition or crossover point decreases with decreasing $\beta_2$. 
For a small $\beta_2$, a steep increase in $d$ with increasing $\alpha$
is also observed.

\subsection{Local spin and charge susceptibilities}

In the bipolaronic state with a large $d$,
it is expected that the local charge fluctuation 
is enhanced together with the local lattice fluctuation,
while the local moment is suppressed.
In fact, from the definition of $d$ in the previous subsection,
we obtain
$\langle{(n_i-\langle {n_i}\rangle)^2}\rangle=2d$
and
$\langle {s_{zi}}^2 \rangle=(1-2d)/4$,
where
$s_{zi}=(c^{\dag}_{i\uparrow}c_{i\uparrow}-
c^{\dag}_{i\downarrow}c_{i\downarrow})/2$.
Then, we examine the local charge and spin susceptibilities
$\chi_{\rm c}^{\rm loc}$ and $\chi_{\rm c}^{\rm loc}$,
defined as
\begin{equation}
 \chi_{\rm c}^{\rm loc}
 =\int_0^{1/T} d\tau \langle n_i(\tau)n_i \rangle,
\end{equation}
and
\begin{equation}
 \chi_{\rm s}^{\rm loc}
  =\int_0^{1/T} d\tau \langle s_{zi}(\tau)s_{zi} \rangle,
\end{equation}
respectively.

\begin{figure}[t]
\begin{center}\leavevmode
\rotatebox{0}{\includegraphics[width=60mm,angle=0]{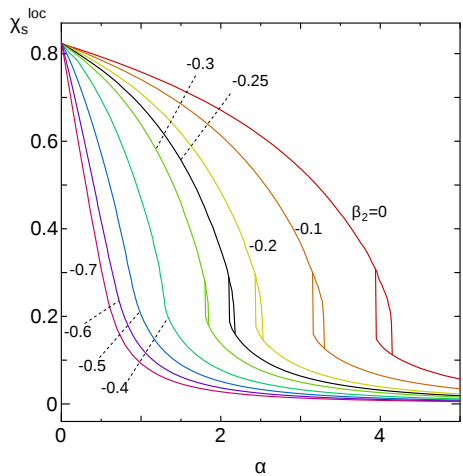}}
\caption{(Color online) 
 Local spin susceptibility ${\chi_s}^\mathrm{loc}$ vs $\alpha$
 for several $\beta_2$ values with $\beta_4=0.02$ and $T=0.0025$.
 \label{fig:alp-chis_t00025}
}
\end{center}
\end{figure}

\begin{figure}[t]
\begin{center}\leavevmode
\rotatebox{0}{\includegraphics[width=64mm,angle=0]{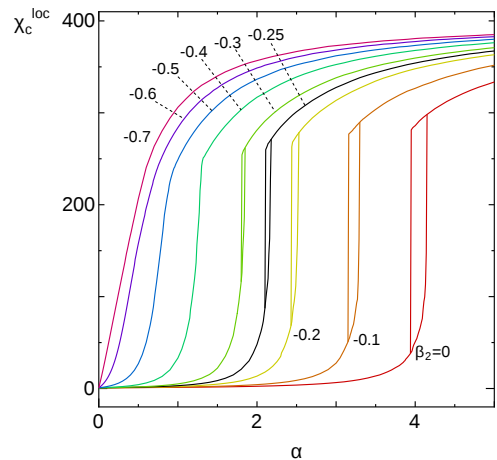}}
\caption{(Color online) 
 Local charge susceptibility ${\chi_{\rm c}}^\mathrm{loc}$ vs $\alpha$
 for several $\beta_2$ values with $\beta_4=0.02$ and $T=0.0025$.
 \label{fig:alp-chic_t00025}
}
\end{center}
\end{figure}

In Figs.~\ref{fig:alp-chis_t00025} and \ref{fig:alp-chic_t00025},
we show ${\chi_{\rm s}}^\mathrm{loc}$ and ${\chi_{\rm c}}^\mathrm{loc}$,
respectively,
for several anharmonic potentials with $T=0.0025$.
With increasing $\alpha$, ${\chi_s}^\mathrm{loc}$ is a decreasing function, 
while ${\chi_c}^\mathrm{loc}$ is an increasing function.
The degree of decrease or increase tendency is enhanced,
when $\beta_2$ is decreased, i.e., the anharmonicity is increased.
The behavior agrees well with our expectation of the local charge and spin
fluctuations, mentioned above.
For the solutions with a sufficiently large $\alpha$, irrespective of $\beta_2$,
${\chi_{\rm s}}^\mathrm{loc}$ asymptotically approaches zero,
while ${\chi_{\rm c}}^\mathrm{loc}$ approaches a finite specific value, $\sim 400$.
The behavior has been obtained in the calculations for
a harmonic phonon.\cite{FuseOno}
Thus, the present result confirms that the states with a large $\alpha$
describe the bipolaronic state, in which charge fluctuations are dominant,
while spin fluctuations are totally suppressed.

\subsection{Renormalization factor}

In this subsection,
we discuss the possibility of the heavy-electron state
due to the strong electron-phonon interaction 
with the anharmonicity.
For this purpose, we estimate the renormalization factor $z$ 
defined by
\begin{equation}
  z^{-1}=1-\left.\frac{d{\rm Re}\Sigma(\omega)}{d\omega} \right|_{\omega=0},
\end{equation}
where $\Sigma(\omega)$ is the electron self-energy with a real frequency $\omega$.
In general, $z$ means the renormalization effect of the conduction electron state.

\begin{figure}[t]
\begin{center}\leavevmode
\rotatebox{0}{\includegraphics[width=70mm,angle=0]{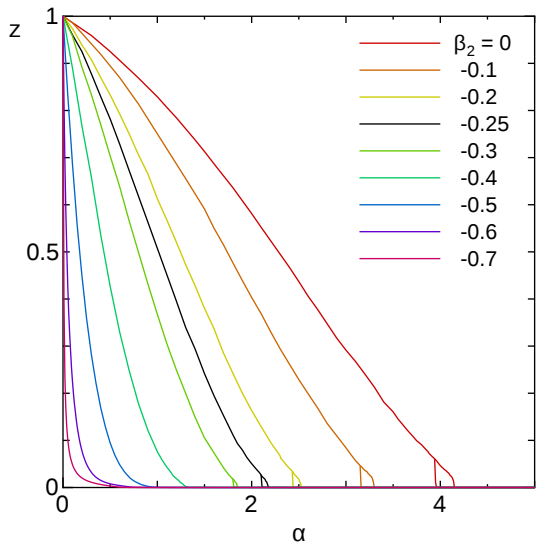}}
\caption{(Color online) 
 Renormalization factor $z$ vs $\alpha$
 for several $\beta_2$ values with $\beta_4=0.02$ and $T=0.0025$.
 \label{fig:alp-Z_t00025}
}
\end{center}
\end{figure}

In Fig.~\ref{fig:alp-Z_t00025},
we plot $z$ as a function of $\alpha$ for several $\beta_2$ values 
at a low temperature $T=0.0025$.
With increasing $\alpha$, irrespective of $\beta_2$ values,
$z$ monotonically decreases and finally goes to zero.
As well as other physical quantities discussed in the previous subsections,
for a weak-anharmonicity cases with $\beta_2 \ge -0.3$,
we find the hysteresis region,
in which large- and small-$z$ solutions coexist at the same $\alpha$.
On the other hand, for strong-anharmonicity cases with $\beta_2 < -0.4$,
the crossover between the two solutions is observed.
The behavior of the decreasing $z$ seems to depend on $\beta_2$.
For the harmonic-like potential with $\beta_2 > -0.25$,
$z$ is a slightly convex-upward function for a small $\alpha$,
while for the double-well potential with $\beta_2 <-0.25$,
it changes to a convex-downward function.
For a smaller $\beta_2$, we find a steeper decrease in $z$
for a small $\alpha$ with a polaronic solution.

Here, we recall that the inverse of $z$ indicates
the mass enhancement of electrons, expressed by $z^{-1}=m^*/m$,
where $m^*$ denotes the effective mass of the conduction electron and
$m$ is the bare electron mass.
Thus, the small-$z$ solution indicates the formation of
a heavy-electron state.
Note that the solution in the limit of $z \rightarrow 0$
indicates the bipolaronic state with a large $\alpha$.
The heavy-electron state should appear in the polaronic phase
in the vicinity of the bipolaronic phase.
In this sense, the case in which the solutions with a small $z$
are widely distributed in the parameter region seems to be
advantageous for the emergence of the heavy-electron state.

For weak-anharmonicity cases such as $\beta_2=0$,
small-$z$ solutions with a possible heavy-electron state are found
only in the vicinity of the two-solution coexistence region.
Thus, there are small possibilities of the heavy-electron state 
for weak-anharmonicity cases.
On the other hand, for strong-anharmonicity case
with crossover solutions such as $\beta_2=-0.5$,
$z$ exhibits a convex-downward behavior in the crossover region.
Namely, small-$z$ solutions are distributed in the relatively wide
region of $\alpha$.
When $\beta_2$ is decreased,
the tail-like structure can be found in the $z$ behavior
in the crossover region,
indicating that the region of small-$z$ solutions becomes wide.
Therefore, the heavy-electron state due to the electron-phonon interaction
is brought about by the large anharmonicity.

\begin{figure}[t]
\begin{center}
\includegraphics[width=70mm,angle=270]{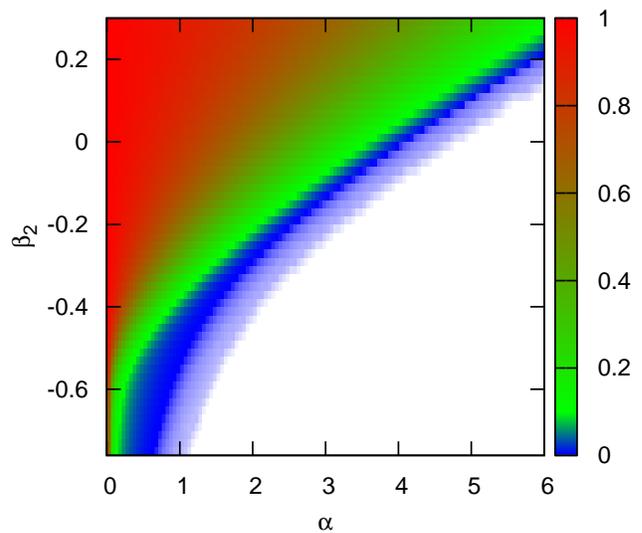}
\end{center}
\caption{(Color online) 
Renormalization factor $z$ on $\alpha$-$\beta_2$ plane for $T=0.0025$.
Depending on the magnitude of $z$, the phase is shown by color gradation 
(contrasting density in the case of black-and-white printing) :
From the left-hand side to the right-hand side of the graph, 
red (dark gray) for $0.1 \lesssim z \le 1$,
green (light gray) for $z$ on the order of $0.1$,
blue (black) for $0.001 \lesssim z \lesssim 0.1$,
and white for $z < 0.001$ (bipolaronic phase).
\label{fig:phasediagram_HF_t00100}
}
\end{figure}

To discuss the heavy-electron state in more detail,
we show the phase diagram concerning $z$ in the $(\alpha$, $\beta_2)$ phase
for $T=0.0025$ in Fig.~\ref{fig:phasediagram_HF_t00100}.
Here, we define that the heavy-electron state is characterized by the solution with
$10 \le z^{-1} \le 1000$ and that the corresponding region is indicated in blue 
(or black, in the case of black-and-white printing).
Note that the polaronic metallic region with a large $z$ is expressed in red (dark gray),
while the polaronic phase with smaller $z$ is shown in green (light gray).
The red (dark gray) and green (light gray) regions are continuously connected and the
polaronic phase with a very small $z$ is defined by the heavy-electron state.
The bipolaronic insulating phase is expressed by the solution
with an extremely small $z$ and is shown in white in the figure.

The green (light gray) region with a smaller $z$ is shown on the left-hand side of
the graph, while the white region with $z\rightarrow 0$ is shown on 
the right-hand side of the figure.
The blue (black) region with the possible heavy-electron state
exists between them.
Note that the narrow blue (black) region with a small $z$ on the order of $1/1000$ 
can be found near the bipolaronic state with $\alpha>\alpha_{\rm c2}$
for large a $\beta_2$,
where $\alpha_{\rm c2}$ is the upper critical value of
the polaron-bipolaron transition.
The white region cannot be regarded as the heavy-electron state
owing to the strong localization in the bipolaronic state.

When the anharmonicity increases with decreasing $\beta_2$,
the small-$z$ region expands in the wider region of $\alpha$
around the two-solution coexistence or crossover region.
When we decrease $\beta_2$,
the expansion of the small-$z$ region is pronounced for
$\beta_2 \lesssim -0.25$, 
where the potential shape is changed from the flat single-well type 
to the double-well type.
The present results suggest the close relation between
the heavy-electron state and the anharmonicity.

\begin{figure}[t]
\begin{center}\leavevmode
\rotatebox{0}{\includegraphics[width=70mm,angle=0]{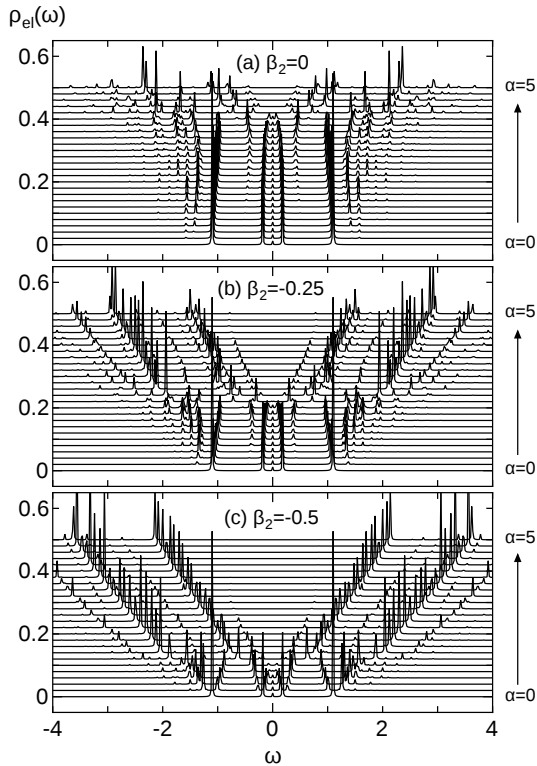}}
\caption{
Electron DOSs $\rho_{\rm el}(\omega)$ for three typical potentials as
(a) harmonic type with $\beta_2=0$,
(b) flat single-well type with $\beta_2=-0.25$, and
(c) double-well type with $\beta_2=-0.5$.
We set $\beta_4=0.02$ at $T=0.0025$.
We show the $\alpha$ dependence from $0$ to $5$ from bottom to top
for each panel.
For each $0.2$ step in the increase in $\alpha$,
the graph is shifted upward by $0.02$.
\label{fig:spct_rhoel}
}
\end{center}
\end{figure}

\subsection{Electron and phonon spectral functions}

Now, we discuss the spectral functions of electrons and phonons.
For this purpose, we evaluate the DOS 
obtained from the imaginary part of Green's function.
The electron DOS $\rho_{\rm el}(\omega)$ and the phonon DOS
$\rho_{\rm ph}(\omega)$ are respectively given by
\begin{equation}
 \rho_{\rm el}(\omega)=-\frac{1}{\pi}{\rm Im}G(\omega+i\eta),
\end{equation}
and
\begin{equation}
 \rho_{\rm ph}(\omega)=-\frac{1}{\pi}{\rm Im}D(\omega+i\eta),
\end{equation}
where $\eta$ is a positive infinitesimal quantity,
$G$ is defined in eq.~(\ref{Gel}),
and the phonon Green's function $D$ is given by
\begin{eqnarray}
 D(i\nu_n) =
  \frac{1}{Z} \sum_{j,\ell}
  \frac{e^{-E_{\ell}/T}-e^{-E_j/T}}
       {i\nu_n +E_{j}-E_{\ell}}
 \left|\langle j | (b_i+b_i^{\dag}) | \ell \rangle\right|^2.
\end{eqnarray}
Here, $\nu_n=2\pi T n$ is the boson Matsubara frequency.
Note that we set $\eta=0.004$ in order to draw
the continuous spectrum function for clear visibility.

In Fig.~\ref{fig:spct_rhoel},
we show the curves for the electron DOS at $T=0.0025$
for three typical potentials (a) $\beta_2=0$, (b) $-0.25$,
and (c) $-0.5$, corresponding to the harmonic,
the flat single-well, and the double-well types,
as shown in Figs.|\ref{fig:potentialshapes}(b)-\ref{fig:potentialshapes}(d), respectively.
For each panel, we show the curves in the order of $\alpha$ from bottom to top.
Note that, in the two-solution coexistence region,
we plot the results for the polaronic metallic solution with a large $z$.

At $\alpha=0$, a set of two large peaks for high energy $\omega>0$
(low energy $\omega<0$) and a small peak at $\omega=0$
can be commonly seen in the three panels.
Note that the bilaterally symmetric graph is
due to the particle-hole symmetry. 
For a small $\alpha$, the structure is almost unchanged.
On the other hand, for a large $\alpha$, the set of
two large peaks in $\omega>0$ ($\omega<0$) moves to
the right-hand side (left-hand side) of the graph
with increasing $\alpha$.
Then, the disappearance of the peak at $\omega=0$ is also observed,
indicating the disappearance of quasi-particle band.
The critical values of the disappearance of the peaks at $\omega=0$
are $\alpha \sim 4$, $2$, and $1$ for $\beta_2=0$, $-0.25$, and $-0.5$,
respectively, which agree well with the $\alpha$ which
the transition or crossover is indicated.

\begin{figure}[t]
\begin{center}\leavevmode
\rotatebox{0}{\includegraphics[width=70mm,angle=0]{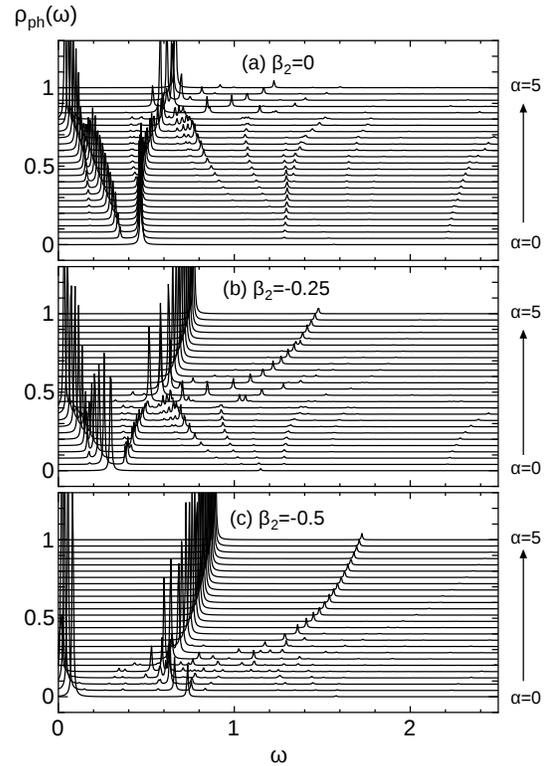}}
\caption{
Phonon DOSs $\rho_{\rm ph}(\omega)$ for three typical potentials as
(a) harmonic type with $\beta_2=0$, 
(b) flat single-well type with $\beta_2=-0.25$, and
(c) double-well type with $\beta_2=-0.5$.
We set $\beta_4=0.02$ at $T=0.0025$. 
We show the $\alpha$ dependence from $0$ to $5$ from bottom to top for each panel.
For each $0.2$ step increase in $\alpha$,
the graph is shifted upward by $0.04$. 
\label{fig:spct_rhoph}
}
\end{center}
\end{figure}

Note here that the renormalization factor $z$ discussed above should
correspond to the decrease in the bandwidth,
which is well known as the polaronic band narrowing effect.
However, it is difficult to observe such a band narrowing effect
for a small $\alpha$ in these graphs,
in spite of the significant change in $z$.
This is seemingly incomprehensible at a glance,
but similar results have already been found
in the harmonic Holstein model.\cite{MEDos,HolstHewson,HHHewson}
According to previous studies,
in the polaronic phase, with increasing effect of electron-phonon interaction,
the central peak at $\omega=0$ becomes narrower and more pronounced
with the corresponding weight $z$,
while the other peaks with total weights of $1-z$ exhibit almost no changes.
The reason why the other peaks do not move at all is
that the Holstein model has few energy distribution corresponding
to the upper and lower Hubbard bands,
which are split upward and downward depending
on the Coulomb interaction.
Thus, the conservation of the DOS for a small $\alpha$ is valid, 
at least within the present calculation on the basis of the exact diagonalization.
As for the reproduction of the central peak behavior, 
it is necessary to improve the precision of the calculation,
for instance, by increasing the cluster size.

Next we move to the phonon spectral function.
In Fig.~\ref{fig:spct_rhoph},
we show $\rho_{\rm ph}(\omega)$ at $T=0.0025$
for three typical potentials with
$\beta_2=0$, $-0.25$, and $-0.5$.
At $\alpha=0$, for each three panel, 
distinct peaks can be seen at $\omega \sim 0.43$, $0.3$, and $0.04$
for $\beta_2=0$, $-0.25$, and $-0.5$, respectively.
These $\omega$ values correspond to quasi-harmonic phonons,
which agree well with the first excitation energies
observed in Fig.~\ref{fig:energylevels}.

With decreasing $\beta_2$, the shift of the peaks
to the lower-energy side is indicated. 
In the polaronic state for small $\alpha$, 
the shifts of the low-energy peaks are significant
with increasing $\alpha$,
implying the softening of phonons and the divergence of
charge susceptibility.
The heavy-electron tendency caused by phonon softening
due to anharmonicity has been discussed
in refs.~\citen{Oshiba1} and \citen{Oshiba2}.
In the polaron-bipolaron transition or crossover region,
the disappearance of the peaks at lower energy can be seen, 
suggesting the marked change in phonon properties.
The critical value of $\alpha$ at which the lowest energy peak disappears
is in good agreement with the transition point or crossover region.
In the bipolaronic state, the second lower peaks are continuously
connected to those in the polaronic state
and they move to the high-energy side with increasing $\alpha$
as well as in the electron spectrum cases.

\begin{figure}[t]
\begin{center}\leavevmode
\rotatebox{0}{\includegraphics[width=70mm,angle=0]{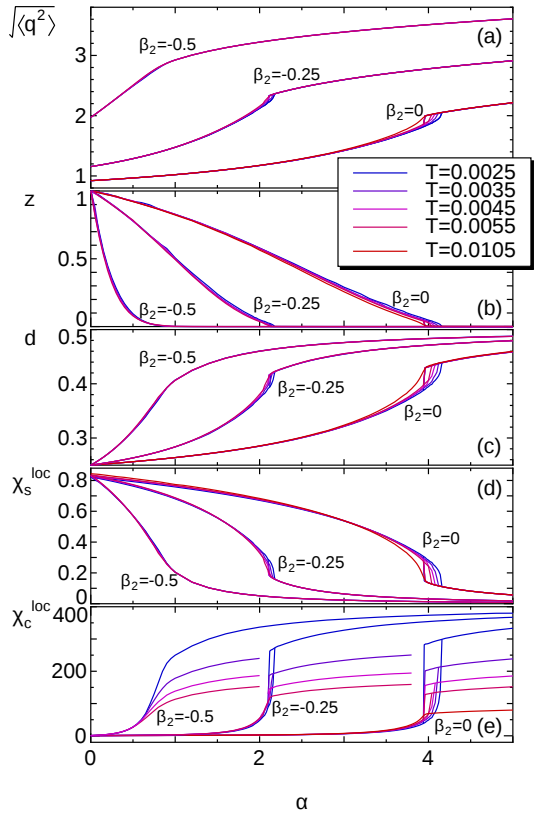}}
\caption{(Color online)
(a) Lattice fluctuation  $\sqrt{\langle{q^2}\rangle}$ vs $\alpha$,
(b) renormalization factor $z$ vs $\alpha$,
(c) double occupancy $d$ vs $\alpha$,
(d) local spin susceptibility $\chi_{\rm s}^\mathrm{loc}$ vs $\alpha$,
and (e) local charge susceptibility $\chi_{\rm c}^\mathrm{loc}$ vs $\alpha$
for $\beta_2=0$, $-0.25$, $-0.5$ with $T=0.0025$, $0.0035$, $0.0045$, and
$0.0055$ with $\beta_4=0.02$. 
In the $\beta_2=0$ case, the graph with the crossover result
at $T=0.0105$ is also indicated. 
Several $\chi_{\rm c}^\mathrm{loc}$ dependences for a large $\alpha$ are omitted 
to avoid overlaps of the graph. 
Note that $\chi_{\rm c}^\mathrm{loc}$ approaches to a finite specific value 
depending on $T$ for a large $\alpha$. 
\label{fig:alp-etc_Tdeps}
}
\end{center}
\end{figure}

\subsection{Temperature dependence of physical quantities}

Thus far, we have discussed several physical quantities
focusing our attention on the fixed temperature with $T=0.0025$. 
In this subsection, we mention the temperature dependence briefly. 
Here, we consider three typical potentials with
$\beta_2=0$, $-0.25$, and $-0.5$ for $\beta_4=0.02$.
In Figs.~\ref{fig:alp-etc_Tdeps},
we show the $\alpha$ dependences of several physical
quantities such as (a) lattice fluctuation $\sqrt{\langle{q^2}\rangle}$,
(b) the renormalization factor $z$,
(c) the double occupancy $d$,
(d) the local spin susceptibility $\chi_{\rm s}^\mathrm{loc}$,
and (e) the local charge susceptibility $\chi_{\rm c}^\mathrm{loc}$
for several temperatures from $T=0.0015$ to $0.0105$.

The marked changes in the physical quantities due to
the difference in temperature 
can be seen only in the vicinity of
the two-solution coexistence or crossover region, 
and the amplitude of the local charge susceptibility.
With increasing temperature, the reduction in the size of the two-solution
coexistence region is indicated.
The edge point for a larger $\alpha$ of
the two-solution coexistence region $\alpha_{\rm c2}$
moves to the left-hand side of the graph.
As for the shift in $\alpha_{\rm c1}$, 
the edge point for a smaller $\alpha$, 
it is relatively small compared with that in $\alpha_{\rm c2}$.
Furthermore, $\alpha_{\rm c1}$ and $\alpha_{\rm c2}$
coincide with each other at high temperatures.
Then, the two-solution coexistence region disappears
and the first-order transition changes to a crossover.
Meanwhile, $\chi_{\rm c}^\mathrm{loc}$ in the bipolaronic solution
for a large $\alpha$ rapidly decreases with increasing $T$.
The temperature dependence of $\chi_{\rm c}^\mathrm{loc}$
in the bipolaronic state exhibits 
the Curie-law behavior $\chi_{\rm c}^\mathrm{loc} \propto 1/T$,\cite{FuseOno}
which is similar to that in the case with the localized spin in the Mott insulator,
where the local spin susceptibility shows
the Curie law behavior $\chi_{\rm s}^\mathrm{loc} \propto 1/T$.
In the harmonic case, the $\chi_{\rm c}^\mathrm{loc}$ in 
the present anharmonic case also follows the Curie law.

\begin{figure}[t]
\begin{center}\leavevmode 
\rotatebox{0}{\includegraphics[width=70mm,angle=0]{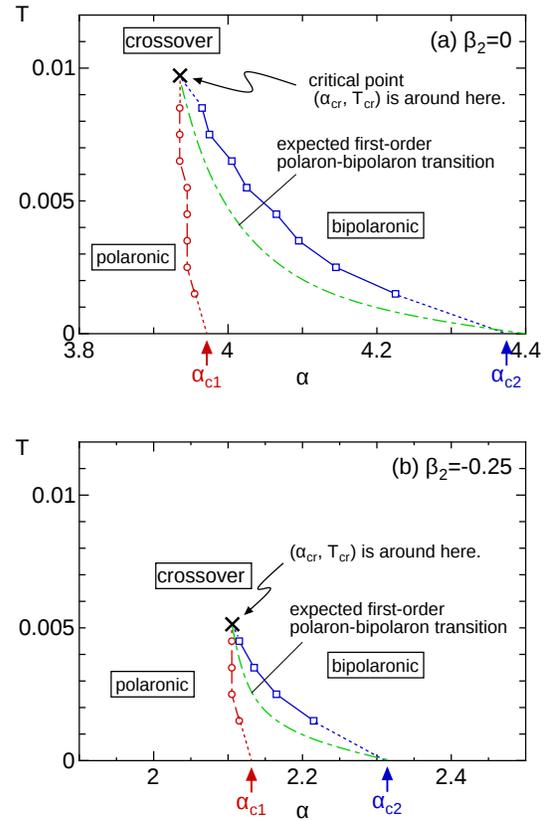}}
\caption{(Color online) 
Phase diagrams on $\alpha$-$T$ plane for
(a) $\beta_2=0$ and (b) $\beta_2=-0.25$ with $\beta_4=0.02$.
In the region $\alpha_{\rm c1}<\alpha<\alpha_{\rm c2}$, 
both the polaronic and bipolaronic solutions coexist.
The dotted-dashed curve shows a possible first-order transition
temperature and the cross indicates its critical point. 
At lower temperatures, we also plot the expected values
of $\alpha_{\rm c1}$ and $\alpha_{\rm c2}$ with a dotted line. 
\label{fig:pd_tw-0000,-0250}
}
\end{center}
\end{figure}

In the strong-anharmonicity case with $\beta_2=-0.5$,
we point out that the crossover behavior is shown at all temperatures,
even below $T=0.0015$ (not shown here).
As indicated in the lowest-energy phonon spectrum
in Fig.~\ref{fig:spct_rhoph}
and the energy levels shown in the bare local phonon potential
in Fig.~\ref{fig:potentialshapes}(d),
the low-temperature crossover is attributed to
the small excitation energy of phonons.
Namely, the amplitude of the excitation energy is considered to
correspond to the energy scale of the polaron-bipolaron transition.
Near the crossover region at high temperatures,
a smooth tail-like behavior with a small $z$ is observed.
For instance, in the graph for $\beta_2=-0.25$,
the tail-like behavior is observed near $\alpha \sim 2.1$
for $T>0.0055$ around the crossover region.
Moreover, for $\beta_2=-0.5$, such behavior can be found
at all the temperature,
indicating the possibility of the heavy-electron state.

\subsection{First-order polaron-bipolaron transition}

In the previous subsection, the two-solution coexistence regions are
indicated for $\beta_2=0$ and $-0.25$ at low temperatures.
Namely, the first-order polaron-bipolaron transition occurs
in the region of $\alpha_{\rm c1}<\alpha<\alpha_{\rm c2}$.
To discuss this transition,
we show the $\alpha$-$T$ phase diagrams for $\beta_2=0$ and $-0.25$
in Figs.~\ref{fig:pd_tw-0000,-0250}(a) and \ref{fig:pd_tw-0000,-0250}(b),
respectively, in which we plot $\alpha_{\rm c1}$ and $\alpha_{\rm c2}$
for several $T$ obtained from the discontinuity
in the lattice fluctuation.

With increasing $T$,
both $\alpha_{\rm c1}$ and $\alpha_{\rm c2}$ show a decreasing tendency,
with the change in the former being moderate,
while that in the latter being pronounced.
We find that $\alpha_{\rm c1}$ and $\alpha_{\rm c2}$ coincide
with each other at a critical point,
denoted by $\alpha_\mathrm{cr}$ or $T_\mathrm{cr}$.
Above this critical temperature,
a smooth crossover is observed instead of a discontinuous
change between the polaronic and bipolaronic solutions.

For $\beta_2=0$ and $-0.25$, the estimated values of $T_\mathrm{cr}$ are 
$\sim 0.01$ and $\sim 0.005$, respectively.
Below $T_\mathrm{cr}$, the first-order polaron-bipolaron transition is expected
to take place similarly to that in the case with the Mott transition.\cite{A.Georges}
Then, we plot a possible first-order transition point in the graph. 
Here, we focus on the largeness of the two-solution coexistence region,
i.e., the region width of $|\alpha_\mathrm{c2}-\alpha_\mathrm{c1}|$
and the critical temperature $T_\mathrm{cr}$ height on the $\alpha$-$T$ plane.
It is found that the area for $\beta_2=-0.25$ is
smaller than that for $\beta_2=0$
in both the $\alpha$ width and the $T_\mathrm{cr}$ height.
Since the reduction in the area can be associated with the energy scale of
the first-order transition,
this result suggests that the enhancement of the anharmonicity
suppresses the first-order polaron-bipolaron transition.

\begin{figure}[t]
\begin{center}
\includegraphics[width=52mm,angle=270]{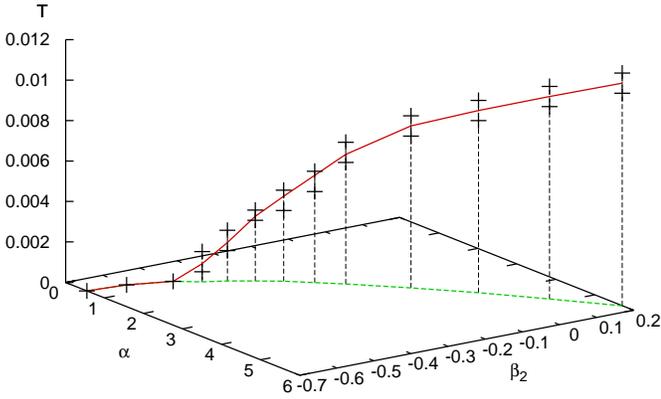}
\caption{(Color online)
First-order polaron-bipolaron transition critical point $\alpha_\mathrm{cr}$, 
$T_\mathrm{cr}$ vs $\beta_2$ for $\beta_4=0.02$.
The crosses indicate the temperatures where the crossover
solution appears on the higher-temperature side and
the coexistence solutions are found
on the lower temperature side. 
The critical points exist
on the bars marked by crosses.
The expected critical points and the projection onto
the $T=0$ plane are shown as solid and dashed curves,
respectively. 
\label{fig:a-b2-T_BP}
}
\end{center}
\end{figure}

Then, we discuss the development of the critical point
of the first-order polaron-bipolaron transition.
In Fig.~\ref{fig:a-b2-T_BP}, we show the $\beta_2$ dependences of 
the critical point $\alpha_\mathrm{cr}$, $T_\mathrm{cr}$.
Note here that $\alpha_\mathrm{cr}$ in the crossover region is
determined from the value at which the curvature of the lattice fluctuation
$\sqrt{\langle q^2\rangle}$ changes.
In the graph, when $\beta_2$ decreases,
both $\alpha_\mathrm{cr}$ and $T_\mathrm{cr}$ decrease.
The decrease in $\alpha_\mathrm{cr}$ indicates the enhancement of
the effective electron-phonon interaction coupled to the anharmonicity.
As for $T_\mathrm{cr}$, we observe a gradual decrease
for a large $\beta_2$, 
while the shape changes to a slightly steeper curve for $\beta_2 \sim -0.25$
and finally approaches zero.
This $T_\mathrm{cr}$ behavior seems to be similar to 
the $\beta_2$ dependence of the first excitation energy of 
phonons in Fig.~\ref{fig:energylevels}. 
Thus, the suppression effect is thought to be caused by the reduction
in the excitation energy of phonons.
For $\beta_2<-0.5$, we find no evidence of the two-solution coexistence region
for the temperature region where we have performed our calculations.
For this reason, we set $T_\mathrm{cr}$ to $0$ in Fig.~\ref{fig:a-b2-T_BP};
however, we do not exclude the possibility of a polaron-bipolaron transition point
with an extremely low transition temperature in the region of
$\beta_2 \lesssim -0.5$,
since a small but finite excitation energy is indicated in the local phonon problem.

\section{Discussion and Summary}

In this paper,
to discuss the effect of the anharmonicity of phonons
on the emergence of the heavy-electron state and the polaron-bipolaron transition,
we have analyzed the anharmonic Holstein model,
which describes the interaction between conduction electrons and 
the atom oscillation in the potential with second- and fourth-order
anharmonic terms,
using the dynamical mean-field theory by the exact diagonalization
method.
We have obtained various physical quantities as functions of
the electron-phonon interaction $\alpha$
for several types of anharmonic potential.

First, we have discussed the effect of anharmonicity
on the first-order polaron-bipolaron transition.
For a weak-anharmonicity case with a large $\beta_2$,
with increasing $\alpha$,
we have observed an increase in lattice fluctuation,
double occupancy, and local charge susceptibility,
but a decrease in the renormalization factor $z$ and
local spin susceptibility.
Moreover, at low temperatures, evidence of the phase transition
from the polaronic state to the bipolaronic state,
accompanied by the changes in the physical quantities,
is found for a strong-coupling region
as the two-solution coexistence region.
When temperature increases, the region becomes narrower
and it finally disappears at a certain critical temperature,
resulting in a smooth crossover between the two states.

The polaron-bipolaron transition behavior has been reported
in a previous study with harmonic phonons \cite{FuseOno}
and there is no qualitative difference between their tendencies.
In a previous study with a phonon frequency $\omega_0/W=0.1$,
the transition temperature is estimated to be on the order of
$1/1000$ of the bandwidth $W$.
Since the transition is dominated by charge fluctuations
and it is a relatively spin-independent phenomenon,
we suggest its similarity to the first-order phase transition $T_p=7.5$K 
in the $\beta$-pyrochlore oxide KOs$_2$O$_6$.
Here, we note that in this case, the polaronic (bipolaronic) state 
corresponds to the $T<T_p$ ($T>T_p$) side. 
On the other hand, when anharmonicity increases with
decreasing $\beta_2$,
we have indicated several changes such as the increase in 
the amplitude of the lattice fluctuation,
the decrease in the critical value of $\alpha$,
and the transition temperature suppression.
Thus, in the anharmonic phonon system,
a strong electron-phonon interaction is $not$ required
for the occurrence of the polaron-bipolaron transition,
even though a lower-temperature environment is necessary.
This is considered to be advantageous for the explanation of the $T_p$
transition.

Then, let us discuss how the anharmonicity contributes
to the heavy-electron state.
In this study, we have discussed the effective mass of electrons
through the renormalization factor $z$ as functions of $\alpha$ and $\beta_2$.
With increasing $\alpha$, $z$ monotonically decreases,
namely, the effective mass increases.
The heavy-electron state with a small-$z$ solution has been found
in the narrow region
of the polaronic phase near the polaron-bipolaron transition.
Although smaller values of $z$ have been indicated in the bipolaronic solutions,
quasi-particles have been destroyed owing to the strong localization tendency.
We have found that the possibility of the heavy-electron formation is large
in the region in which the small-$z$ solutions are widely distributed.
In this sense, the heavy-electron state can be found in the region
where the crossover is indicated.
The crossover behavior can be seen even in the weak-anharmonicity cases with a large
$\beta_2$ at high temperatures,
although at low temperatures, it becomes a polaron-bipolaron transition
with a narrow small-$z$ region in the polaronic phase.
Thus, the crossover region with a high possibility of exhibiting the heavy-electron state is only
found in the strong-anharmonicity case with small $\beta_2$.

Throughout this paper, we have restricted ourselves to the case with the normal state 
and have not considered the possible phase transition to the charge-density-wave (CDW) state, 
which is realized in the Holstein model. 
Within the DMFT, the transition temperature $T_\mathrm{CDW}$ to the CDW state in the Holstein model was 
found to be $O(1/100)$ of the bandwidth.\cite{FreericksCDW1} 
As the critical temperature $T_\mathrm{cr}$ of the polaron-bipolaron transition 
obtained in the present study is $O(1/1000)$ of the bandwidth 
and is much smaller than $T_\mathrm{CDW}$, 
it seems to be impossible to observe the polaron-bipolaron transition 
in the normal state above $T_\mathrm{CDW}$. 
The effect of the frustration, however, is expected to largely suppress $T_\mathrm{CDW}$, 
while keeping $T_\mathrm{cr}$ almost constant as the polaron-bipolaron transition 
is exclusively determined by the local DOS, resulting in $T_\mathrm{CDW}<T_\mathrm{cr}$. 
The effect of the anharmonicity $\beta_4$ was also found to suppress $T_\mathrm{CDW}$ except 
in a small $\beta_4$ regime.\cite{FreericksCDW2} 
In addition, in the presence of the Coulomb interaction between electrons, 
the polaron-bipolaron transition in the Hubbard-Holstein model was found to be realized 
for a large-$g$ regime, where $T_\mathrm{cr}$ markedly increases owing to the Coulomb interaction 
and becomes $O(1/100)$ of the bandwidth,\cite{FuseOno3} 
while $T_\mathrm{CDW}$ is not markedly increases but rather reduced.\cite{FreericksCDW3} 
Therefore, we may expect that the polaron-bipolaron transition will be 
observed in the normal state above $T_\mathrm{CDW}$ in the case with the frustration 
and/or Coulomb interaction. 
Explicit calculation including the effects of the frustration 
and the Coulomb interaction together with the effect of the doping\cite{Capone,MitsumotoJMMM} 
will be an important future problem to address.

In summary, we have investigated the half-filled anharmonic Holstein model
using the dynamical mean-field theory in combination with
the exact diagonalization method. 
We have found that, for the weak-anharmonicity case,
the first-order polaronic-bipolaronic phase transition takes place at a critical value
of the electron-phonon coupling $\alpha$, at which each physical
quantity shows discontinuity.
When the anharmonicity is enhanced, the polaron-bipolaron transition temperature
is reduced and the critical value of $\alpha$ decreases.
For a strong-anharmonicity case, 
the polaron-bipolaron transition eventually changes to a crossover,
in which a heavy-electron state with a large effective mass 
is realized owing to the effect of anharmonic phonons.

\begin{acknowledgment}

This work has been supported by a Grant-in-Aid 
for Scientific Research on Innovative Areas ``Heavy Electrons''
(Nos. 20102008 and 23102709) for the Ministry of Education, Culture,
Sports, Science, and Technology, Japan. 
This work has also been supported by a Grant-in-Aid 
for Specially Promoted Research (No. 18002008) and 
for Scientific Research (C) (No. 23540443) in part. 

\end{acknowledgment}

\end{document}